\documentclass[journal]{IEEEtran}
\usepackage{xcolor}

\ifCLASSINFOpdf
\else
\fi
\usepackage{mathtools}
\usepackage{xcolor}

\usepackage{graphicx}
\usepackage{caption}
\usepackage{subcaption}

\usepackage{graphicx}
\usepackage{flexisym}
\usepackage{tikz}
\usetikzlibrary{shapes,arrows}
\usepackage{mathtools}
\usepackage{adjustbox}

\usepackage{lipsum}

\usepackage{atbegshi}
\AtBeginDocument{\AtBeginShipoutNext{\AtBeginShipoutDiscard}}

\begin{document}
%
\title{Deep Neural Networks Meet\\ CSI-Based Authentication }
%
%
%
\author{\IEEEauthorblockN{\normalsize
  Amirhossein Yazdani Abyaneh, Ali Hosein Gharari Foumani and Vahid Pourahmadi\\}
  \IEEEauthorblockA {Department of Electrical Engineering, Amirkabir University of Technology, Tehran, Iran}
  }
  \thanks{Corresponding Author: Vahid Pourahmadi, v.pourahmadi@aut.ac.ir}

\maketitle

\begin{abstract}
The first step of a secure communication is authenticating legible users and detecting the malicious ones. In the last recent years some promising schemes proposed using wireless medium network's features, in particular channel state information (CSI) as a means for authentication. These schemes mainly compare user's previous CSI with the new received CSI to determine if the user is in fact what it is claiming to be. 
Despite high accuracy, these approaches lack the stability in authentication when the users rotate in their positions. This is due to significant change in CSI when a user rotates which mislead the authenticator when it compares the new CSI with the previous  ones. Our approach presents a way of extracting features from raw CSI measurements which are stable towards rotation. We extract these features by the means of deep neural network. We also present a scenario in which users can be {efficiently} authenticated while they are at certain locations in an environment (even if they rotate); and, they will be rejected if they change their location. Also experimental results are presented to show the performance of the proposed scheme.
\end{abstract}


%
\IEEEpeerreviewmaketitle

\section{Introduction}
%
%
%
%
\IEEEPARstart The communication network's architecture  has been divided by layers according to the open systems interconnection model (OSI) \cite{ref:osi}. This approach helps network application developer, as it assigns different tasks to different layers. For instance 
network services are offered by higher layers while having lower layers providing the communication links. In this paradigm encryption or authentication are usually assigned to higher layers \cite{ref:broadcasting}. Most of the conventional  security protocols are based on complexity of solving of an inverse problem and relying on attacker's  limited computational capability. The power of different protocols, thus, depends on how time consuming is to solve that problem. The mechanism and the overhead of  key management and distribution are also among the main features determining the power of a security protocol.

For wireless communication links, instead of using higher layer  protocols, several techniques have been proposed that try to ensure security right at the Physical Layer of the OSI layering, \cite{ref:broadcasting}, \cite{ref:stan3}.These methods are designed to exploit the features of the wireless medium itself and usually use channel information 
to propose  powerful and practical security protocols for encryption and authentication.



Authentications based on non-cryptographic methods have been proposed by \cite{ref:stan6}, \cite{ref:stan23}, \cite{ref:eduard} and \cite{ref:stan}. In \cite{ref:stan6} and \cite{ref:stan23} the systems use RSS or Channel Impulse Response (CIR) to build fingerprints of wireless channel to be further used for authentication. The unique spatial properties of CIR and RSS due to path loss and multi-path effects of the wireless channel privileges us to use them as authenticating parameters. 
In  \cite{ref:eduard} and \cite{ref:stan}, authentication is based on building legitimate profiles for legitimate users based on the users' Channel State Information (CSI). The CSI for these schemes are taken from multiple subcarriers and it is assumed that orthogonal frequency-division multiplexing (OFDM) is used as the transmission methodology. 
In \cite{ref:eduard}, the absolute values of CSI are used for authentication, the method is based on General likelihood ratio tests. To authenticate mobile users, in \cite{ref:stan}, a correlation based method is proposed where the correlation is computed between the new CSIs of mobile users and the CSI collected from the user in the past time slot(s). In the mobile case, the paper presumes the change in CSI to be little, so they used a correlation comparison between adjacent CSIs received.

Considering the low overhead and protocol simplicity, the results of above Physical-layer based authentication schemes are very promising. The one important issue is that correlation between the user's CSIs can be very different if the user rotates, and its variations can be much more significant than the variations in CSI when user moving along a straight path. Due to this rapid change, the correlation method, as we described some of them, would categorized a legitimate user who has rotated as a malicious attacker and  consequently  it  will terminate user's connection with the user.
 
 Motivated by this problem, in this paper, we introduce a method for extracting a set of features from the raw CSI measurements which are stable towards rotation. We derive these features (that are rotation-invariant) by using deep neural networks ( both dense and  convolutional layers)  and implementing a few deep network tricks to help convergence of the network in training phase. 
 As for the experiment, 
we consider an office where users can be authenticated when they are in certain locations of the office. Users in these locations will be able to access the network (regardless of the orientation of their device) and the system will not authenticate them if they are not in these locations.

The structure of the paper is as follows. In Section II we provide the preliminary materials. We explain our system model which is an application of our approach  in Section III. The deep-net for feature selection is presented in Section IV, which will be further used in the authentication framework introduced in section V. Experimental results will be discussed  in Section VI, and Section VII concludes the paper.

\section{Preliminary}
\subsection{Channel State Information}
 
When a signal is sent from a wireless source, it will pass through the channel before it gets to destination. One way to look at the channel is to look at the received signal strength (RSS), which to some extent explain how the channel attenuated the signal but it does not give more accurate information.  

In OFDM data is encoded on multiple subcarrier frequencies. The data is divided into multiple streams which are then coded and modulated respectively on different subcarriers. The subcarrier frequencies are chosen so that they will be orthogonal which results in a minimized interference. For instance, in OFDM employed by 802.11 a/g/n physical layer, a wideband channel with 20 or 40 MHz uses 54 or 108 subcarriers for data transmission, each subcarrier being considered as a narrowband channel. The effect of channel can be see as the total effect of channel on each of the subcarriers. The channel state information (CSI) represent the effect of channel on each of these subcarriers.  Considering $\mathbf{x}_i$ and $\mathbf{y}_i$ as the transmitted and received signal at each subcarrier. For subcarrier $i$ we have: 
\begin{equation} \label{eq:1}
\mathbf{y}_i= \mathbf{H}_i\mathbf{x}_i + \mathbf{n}_i
\end{equation}
where $i$ denotes the number of subcarrier, $\mathbf{n}_i$ is the additive white Gaussian noise and $\mathbf{H}_i$ represents the channel gain at the $i^{th}$ subcarrier. The set of $\mathbf{H}_i$ for all subaccariers is what we refer to as the channel CSI and it captures lots of information about the multipath structure of the environment.  

Commercial NICs (Network Interface controllers) usually do not provide upper network layers with CSI data which is derived at the physical layer. Thanks to \cite{ref:csitool}, with few modifications to Intel 5300 i NIC, CSI data is also reported to the higher layers. Intel Wi-Fi 5300 i NIC implements an OFDM system with 52 subcarriers, which 30 of them will give us CSI.

\subsection{Neural Networks}
\begin{center}
\begin{figure}
  \includegraphics[scale=0.5, width=\linewidth, height =50mm]{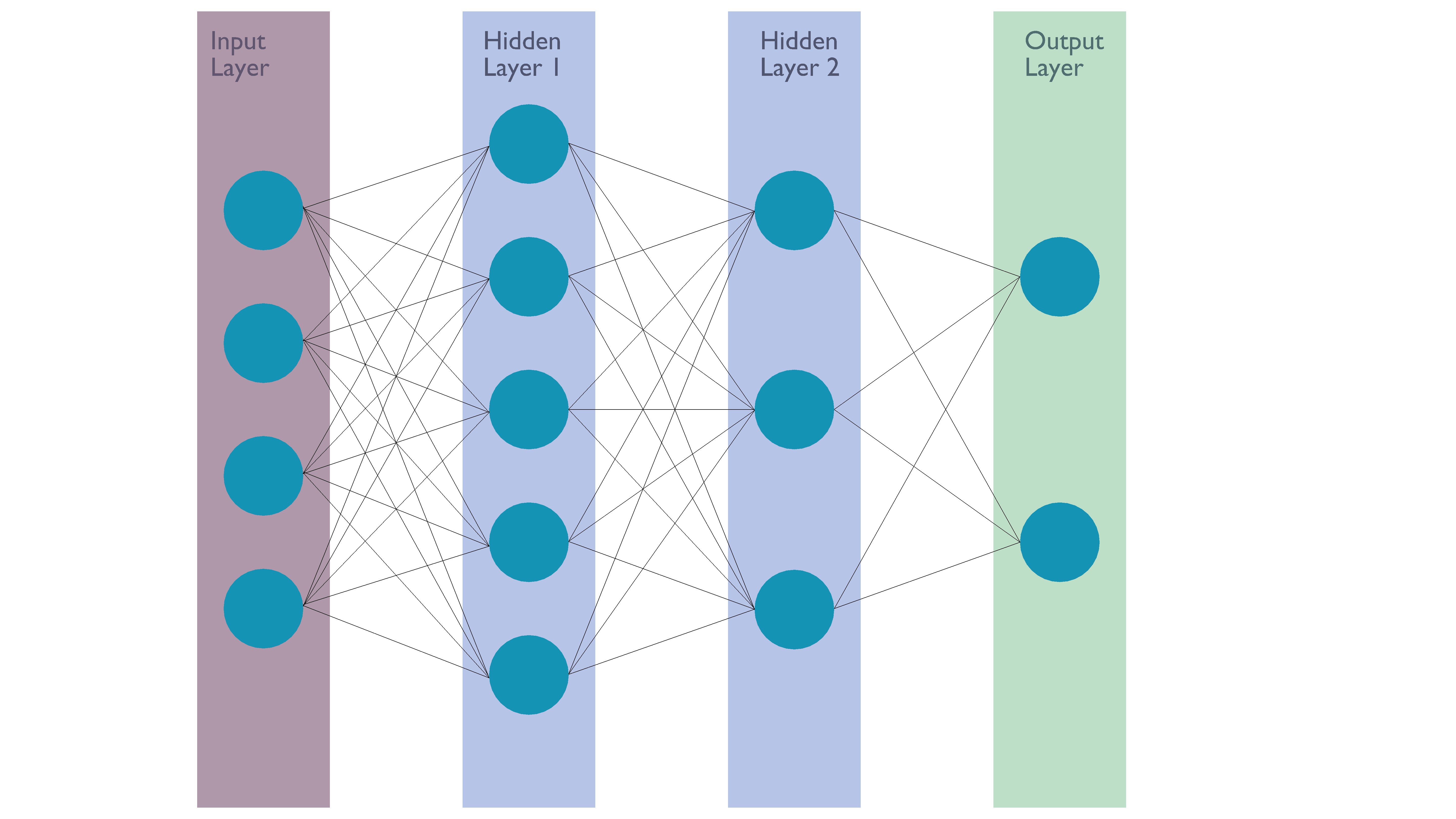}
  \caption{A feed-forward neural network consists of neurons (circles) which are connected to each other by links. Each dense NN should have one input and one output layer; however it could be consisted of none (shallow ANN) or a number of hidden layers (deep ANN) located between the input layer and the output layer. }
  \label{fig:aval}
\end{figure}
\end{center}

Artificial Neural Networks (ANN) are computing systems which are inspired by biological neural networks. These systems learn tasks by examples. ANNs are based on collection of connected nodes called artificial neurons. Each connection between neurons can transmit signals, neurons can process signals and then next layer's neurons connected to it  \cite{ref:neurall}. In ANN implementations, signals are real numbers. The output of each neuron is computed by passing the sum of inputs of that neuron through a nonlinear function. The connections between neurons are called edges. Each edge has a weight which is modified through the process of ANN being trained (ANN learning) \cite{ref:neurall}. You could see a dense neural network (feed-forward neural net) in figure \ref{fig:aval}.

\begin{center}
\begin{figure}
  \includegraphics[scale=0.5, width=\linewidth]{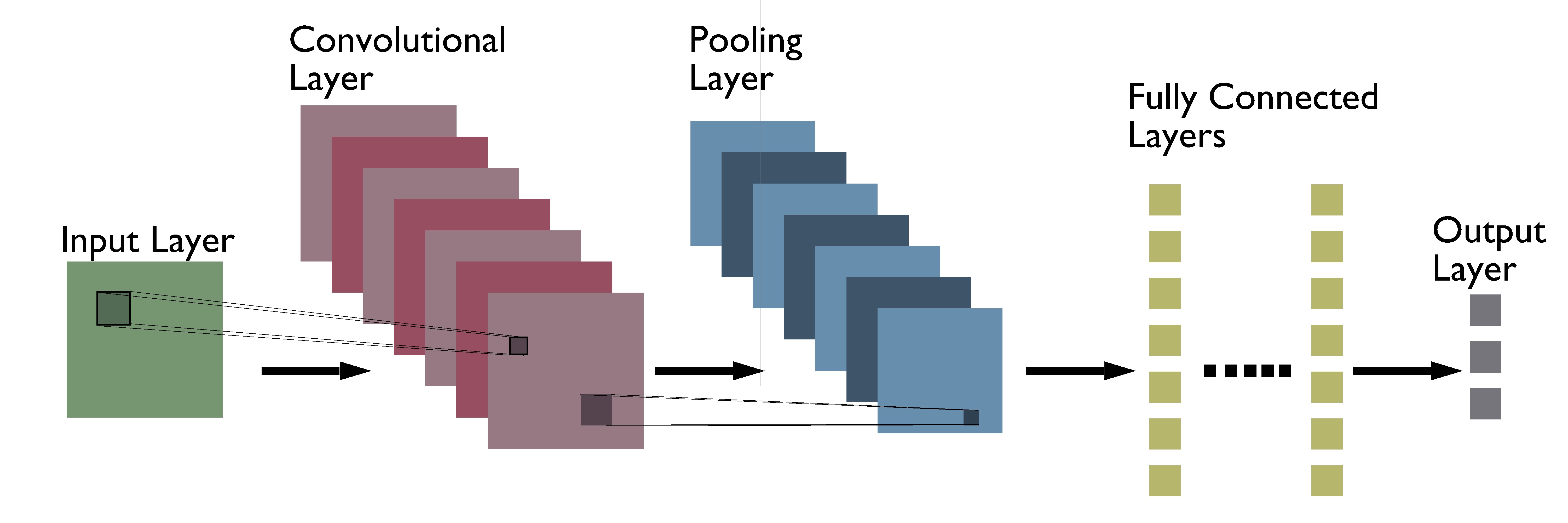}
  \caption{Typical CNN architecture}
  \label{fig:blendconv}
\end{figure}
\end{center}
\subsection{Convolutional Neural Network (CNN) }
Convolutional Neural Networks (ConvNets) are very much similar to ordinary neural networks. They assume that the inputs to the network are image-like (the data which are spatially close are somehow correlated). This property leads to specific structure of network, which is a more efficient forward function being implemented and reduced number of parameters in the network. CNN uses local information of the data by having a multi-dimensional filter which is used for convolution operation of convolutional layers.


 Neurons in the layers of ConvNets are arranged in 3 dimensions (filters) : width, height and depth (depth referring to the third dimension of activation function). Neurons in a specific layer will only be connected to a small region of the previous layer, instead of all neurons in a fully-connected manner (ordinary neural networks) \cite{ref:git}.
  Most of CNNs are  consisted of two distinct parts, the first part is the feature extraction network which is consisted of convolutional and pooling layers. Convolutional layers get an image as an input and produce feature maps, the number of feature maps are equal to number of filters the layer has. Pooling layers merge adjacent pixels based on specific mathematical expressions which the pooling layer is based on. The second part is the classifier network, this network is a fully connected one. Figure \ref{fig:blendconv} shows a complete representation of a convolutional neural  network  consisted of both feature extraction and classification parts.



\section{Authentication Framework} \label{sec:model}

To explain the framework, consider an environment like a large office. We require our model to let the users get positively authenticated if they want to communicate to the network while they are  at $M$ specific locations of this environment, and rejected if they are not located in the specific pre-authenticated locations. We also want give the user the flexibility of rotation (putting their device in any orientation) while they are at that "M" authenticated locations. 

Authentication based on CSI seems a very good fit to the aforementioned application, the issue however is that conventional CSI based authentication schemes cannot handle the rotation of users as the CSI of the user will significantly and rapidly change if they rotate even if they do not move. Therefore, the correlation between two CSI values (before and after rotation) are small which mislead schemes like \cite{ref:stan} and may result in termination of the user after rotation.

Our system tries to compensate this issue by training the system in a way that rotation won't affect our decision. More specifically, we extract some features which are invariant to rotation from the raw CSI measurements . These features can be used as physical layer authenticating parameters. 

Through  the following sections, we first describe the architecture of the deep network that is used for extraction of stable features from the raw CSI measurements (LocNet is used to refer to this network). Next we will discuss how the proposed LocNet is used for the location-based authentication application.

\section{LocNet Architecture}

LocNet takes two inputs and decides whether or not its inputs are from the same location. The network is trained such that it can differentiate between the CSIs coming from different locations.

\begin{center}
\begin{figure}
  \includegraphics[scale=0.5, width=\linewidth, height=65mm]{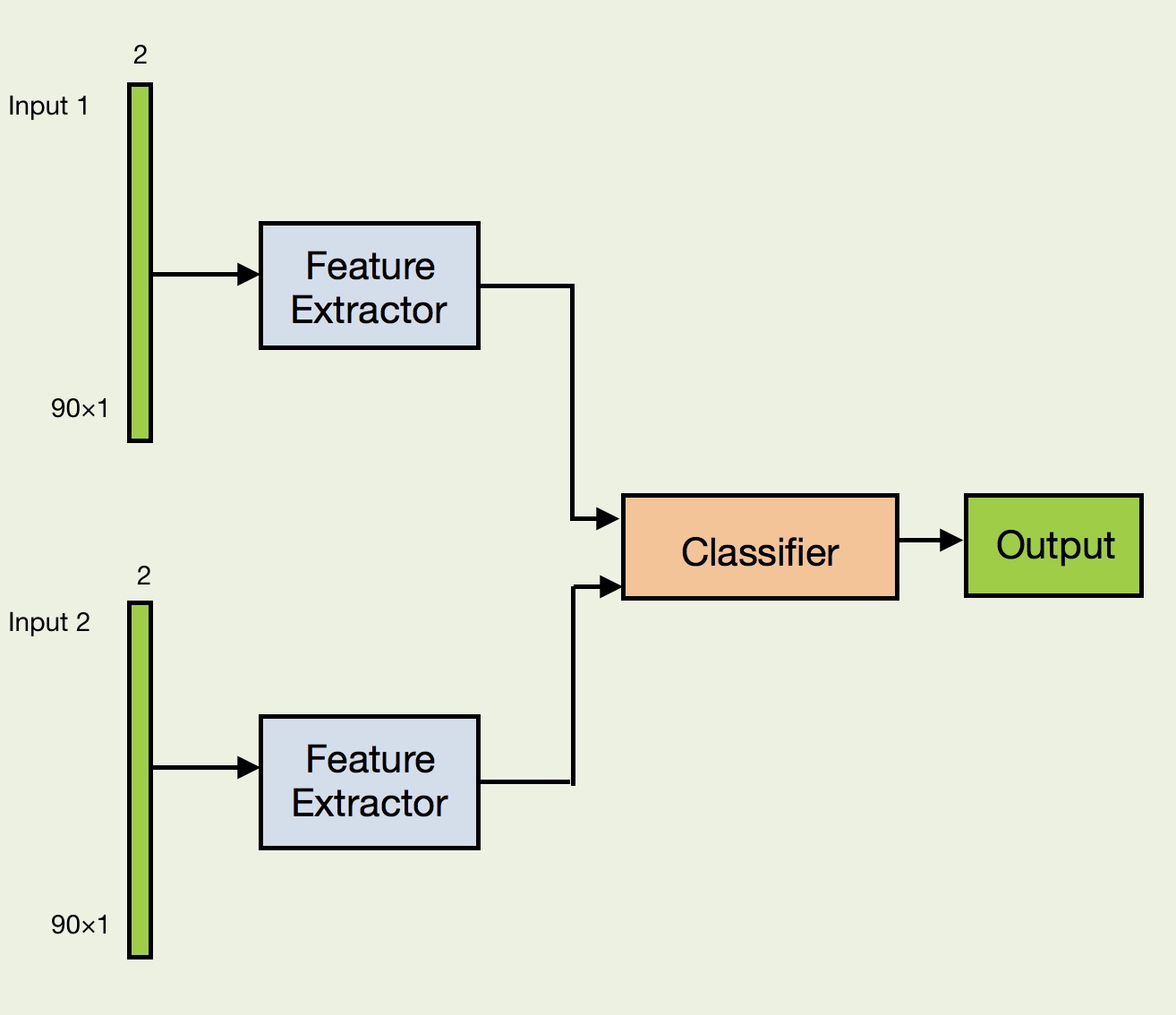}
  \caption{Architecture of LocNet}
  \label{fig:ANN}
\end{figure}
\end{center}

 LocNet  consist of two feature extractors and a classifier (figure \ref{fig:ANN}). In all of our experiments, inputs were 90-dimensional complex vectors (receiver had 3 antennas and we had 30 CSI data (30 subcarrier measurement) for each antenna). Then, inputs go through two feature extractors. The outputs of feature extractors only depend on the location which the CSIs are measured (constant towards rotation). Afterwards, the classifier decides whether or not its inputs are from the same location or not.


\subsection{Feature Extractor}
Our feature extractor is the most important  component of the LocNet. After getting the network trained, if we consider CSI coming from a specific location as an input for our feature extractor, the output will be a 10-dimensional vector which is stable towards rotation. By giving these CSIs to our simple classifier we get a promising results saying that these CSIs are from the same location. 

\begin{center}
\begin{figure}
  \includegraphics[scale=0.5, width=\linewidth,height =50mm]{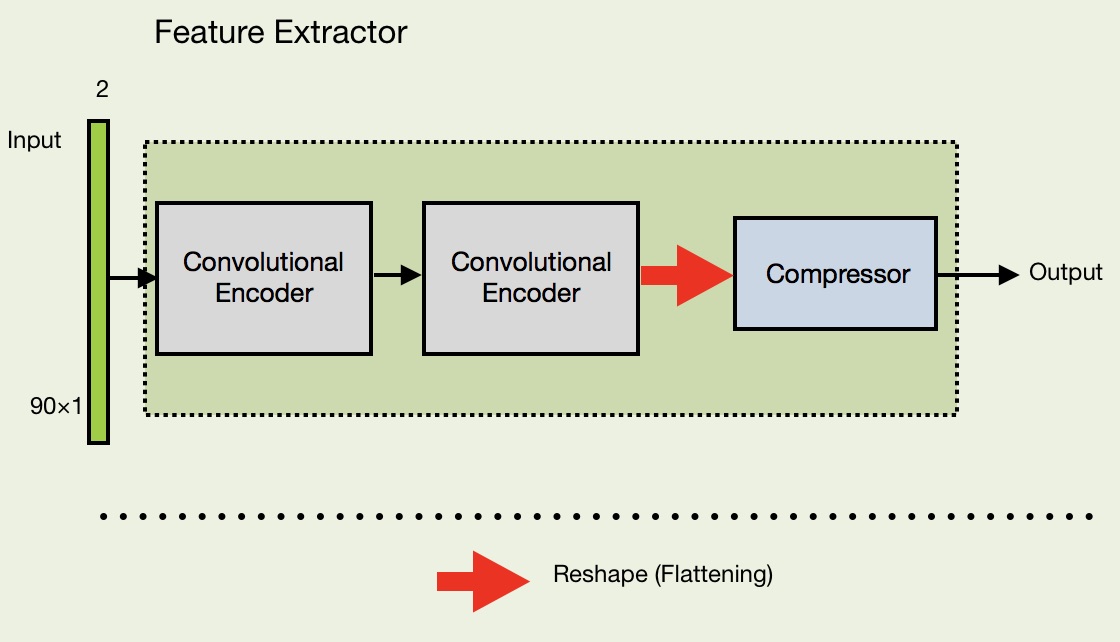}
  \caption{Feature Extractor's architecture  }
  \label{fig:featureextractor}
\end{figure}
\end{center}

The model consists of two identical convolutional networks (ConvEncoders), similar to {\color{blue} \cite{ref:Massive}}, and then it gets flattened and passes through a one feed-forward (dense) network (Compressor). 
The architecture of our feature extractor is depicted in fig \ref{fig:featureextractor}.

\subsubsection{Architecture of the ConvEncoders}
\begin{center}
\begin{figure}
  \includegraphics[scale=0.5, width=\linewidth,height =40mm]{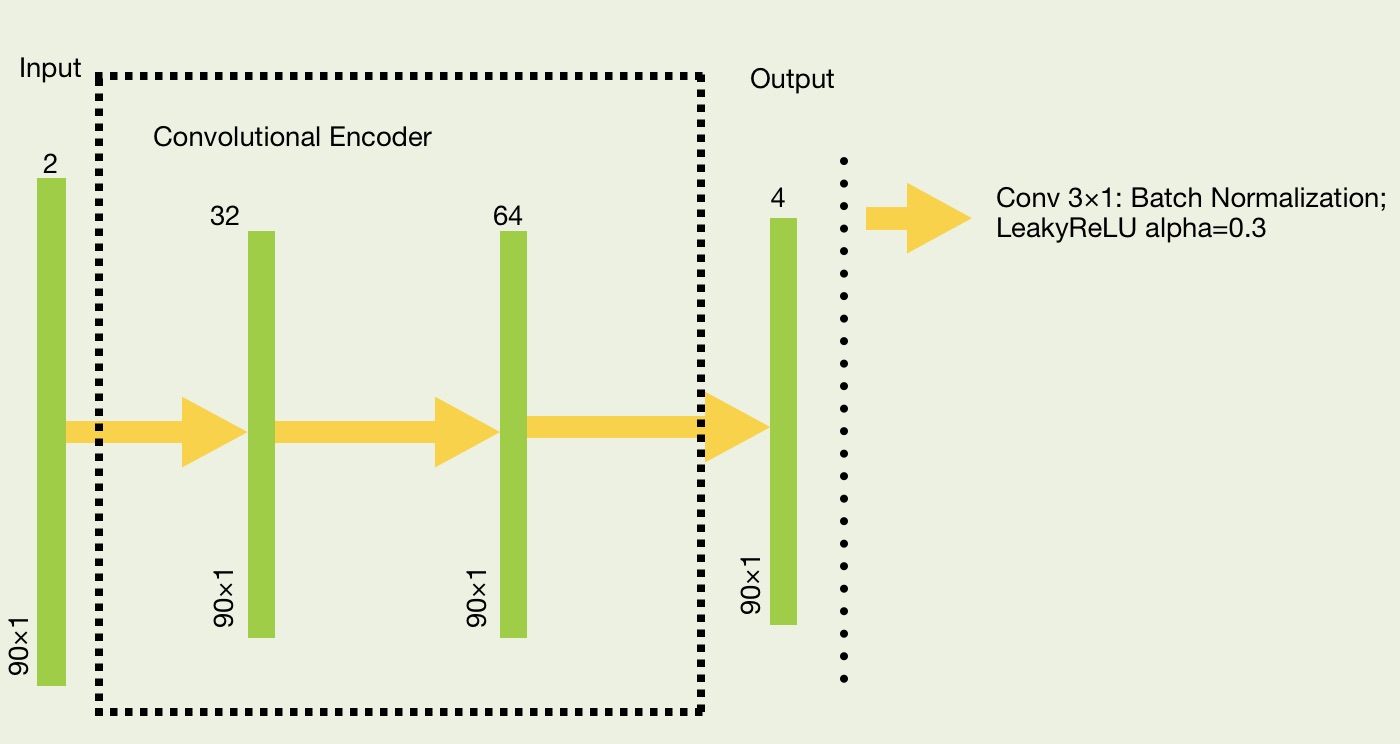}
  \caption{ConvEncoders Architecture  }
  \label{fig:convencoder}
\end{figure}
\end{center}

The feature extractor has two ConvEncoders blocks connected together. The structure of each ConvEncoder is shown in figure \ref{fig:convencoder}.
The input of size $90\times1\times2$ goes through a convolutional layer with 32-filters  (each filter having a size of $3\times1$), the layers uses Batch Normalization  \cite{ref:BNA} and have  Leaky ReLU  activation functions with an $\alpha=0.3$ \cite{ref:ReLU}. The result, then, goes through two other convolutional layers with 64 and 4 filters, respectively; which have the same specifications as the first layer. Each convolutional layer  keeps it output size same as the input.

\subsubsection{Compressor}
\begin{center}
\begin{figure}
  \includegraphics[scale=0.5, width=\linewidth,height =40mm]{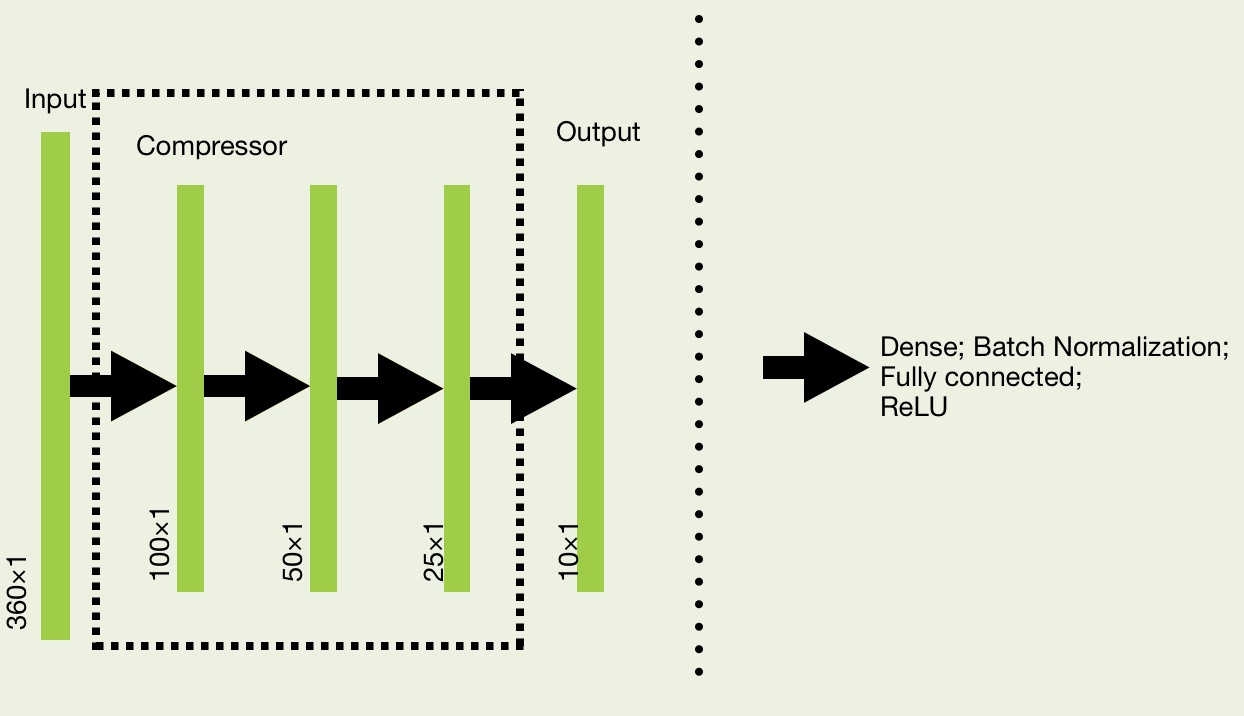}
  \caption{Compressor's Architecture  }
  \label{fig:compressor}
\end{figure}
\end{center}

The Feature Extractor module uses one block of feed-forward dense layers which we call Compressor because its output layer gives us a 10-dimensional vector (from dimensions of 360 to 10). Compressor's architecture is shown in figure \ref{fig:compressor}. First the output of the second ConvEncoder (of size $90\times1\times4$) gets flattened; hence, we have an input vector of size $360\times1$ and its passes through fully connected layers having 100, 50, 25 and 10 neurons respectively. We use Batch Normalization algorithm and ReLU function as the activation function for all layers (except the last one).

\subsection{Classifier}

Passing through the two feature extraction units, the two input CSI vectors turn into two vectors of size 10 which mainly captures the specifications of each location and has less dependency on the orientation of the transmitter.  
A classifier network is then responsible to determine if the two inputs are from a same location or not.

\begin{center}
\begin{figure}
  \includegraphics[scale=0.5, width=\linewidth,height =40mm]{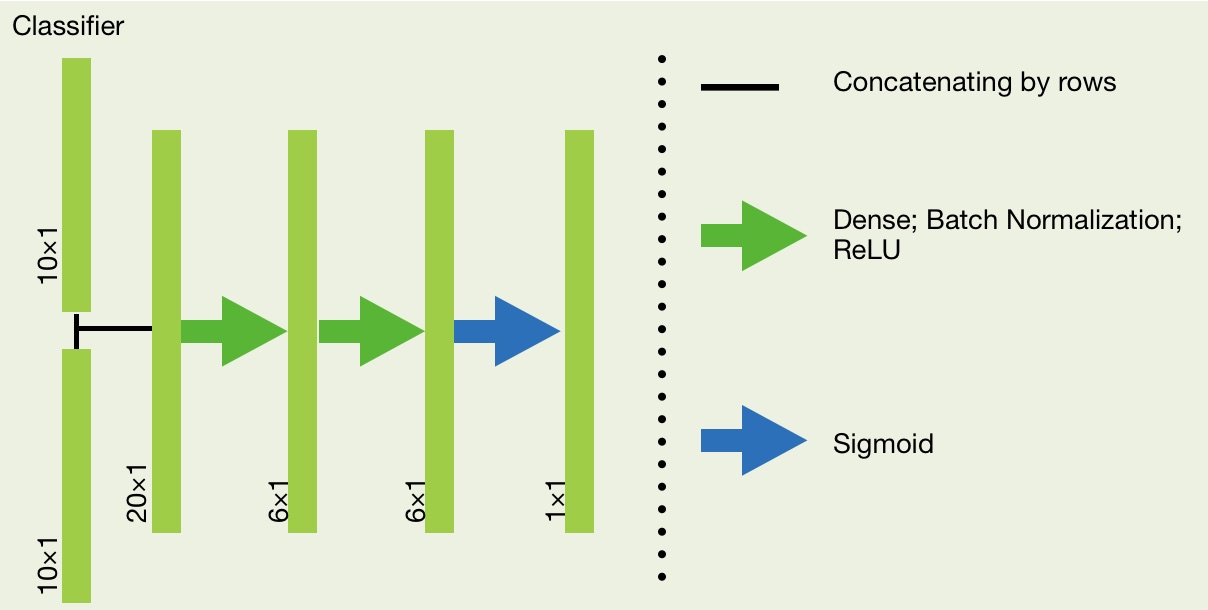}
  \caption{Classifier's Architecture  }
  \label{fig:classifier}
\end{figure}
\end{center}

The classifier's architecture is shown in figure \ref{fig:classifier}. First the two inputs get concatenated so we have a input of size 20. This input goes through three fully connected layers having 6, 6 and 1 neurons, respectively. The activation function for all layers is ReLU except the last one which is the Sigmoid function. The output of the Sigmoid function is then rounded to give us the 0 and 1 output.







\section{Location Based Authentication Framework}

In this section we will discuss how the proposed LocNet can be used to implement the authentication framework discussed in \ref{sec:model}. First, training procedure is described in section \ref{ssec:train}. Then, in 
section \ref{ssec:auth}, we will discuss the procedure for user authentication.

\subsection{LocNet Training} \label{ssec:train}

Considering $M$ locations in the environment as authenticated locations, the aim of LocNet is to find out if its two inputs are coming from the same location or not. For training, we first collect $L$ CSI measurements at each of the $M$ desired locations. The recorded CSIs are gathered while rotating the device on those locations. The training dataset is constructed as below: 


\begin{itemize}
\item[1]
We randomly select one location out of the $M$ locations.
\item[2]
We select one CSI randomly from the set of "L" CSI collected for that location. 
We consider this CSI as the first input, $C(u_1)$ where $1\leq u_1\leq M$.
\item[3]
Again, we randomly select a location from the $M$ locations.
\item[4]
We select one CSI randomly from the location of the third step's selected.
We consider this CSI as the second input, $C(u_2)$ where $1\leq u_2\leq M$.
\item[5]
The training data will be the triplets of $\langle C(u_1), C(u_2), Output \rangle$ where the output is:
\begin{equation} \label{eq:6}
  \mbox{$Output$} =
  \begin{dcases*} 
  0 \quad ~~\quad u_1 =u_2 \\ 
  1 \quad ~~\quad u_1 \neq u_2 
  \end{dcases*}, 
\end{equation}
and shows if the two selected CSI are from one location or not. 

\item[6]
Steps 1-5 should be repeated until it generate $N$ training data. 

\end{itemize}

Binary cross-entropy between the output of the network and the correct output is selected as the network loss function, and Nadam optimizer \cite{ref:NADAM} is selected as the optimization scheme.

\subsection{Authentication Procedure} \label{ssec:auth}

After training, LocNet has the required information for all $M$ locations. Now, when a user wants to establish connection, it should prove that it is in one of the $M$ locations. To this end, the user has to measure CSI $p$ times and send this information to the authentication server.



The authentication server then starts from the first location and generate $K$ test data of the form $\langle T_1,T_2 \rangle$ where $T_1$ is a CSI vector randomly selected from the set of $L$ CSIs has been previously collected for the first location, and $T_2$ is a CSI vector randomly selected from the set of $p$ CSIs reported by the user under test. 

The server feeds LocNet with these set of $K$ test samples. For each test sample, the output will be a zero or one, zero means the network believes that the user is not in the first location, and one means the network believe it is there. We will have $K$ responses from the LocNet, we use $r$ to denote the number of test cases that are equal to one. The authentication server then declares whether or not the user is at first location, if it passes the majority test, i.e., $\alpha=\frac{r}{ K} >\zeta$. $\zeta$ is a threshold parameter that can be used to adjust the required confidence level before authorizing the user to access the network. If $\zeta$ has a large value the authentication server only authorizes user if most of the test data agree that the user is at the first location, and if $zeta$ is close to 0.5, it authorizes the user if more that half of the test cases agree on that.

\begin{figure}
  \centering
  \includegraphics[scale=0.25]{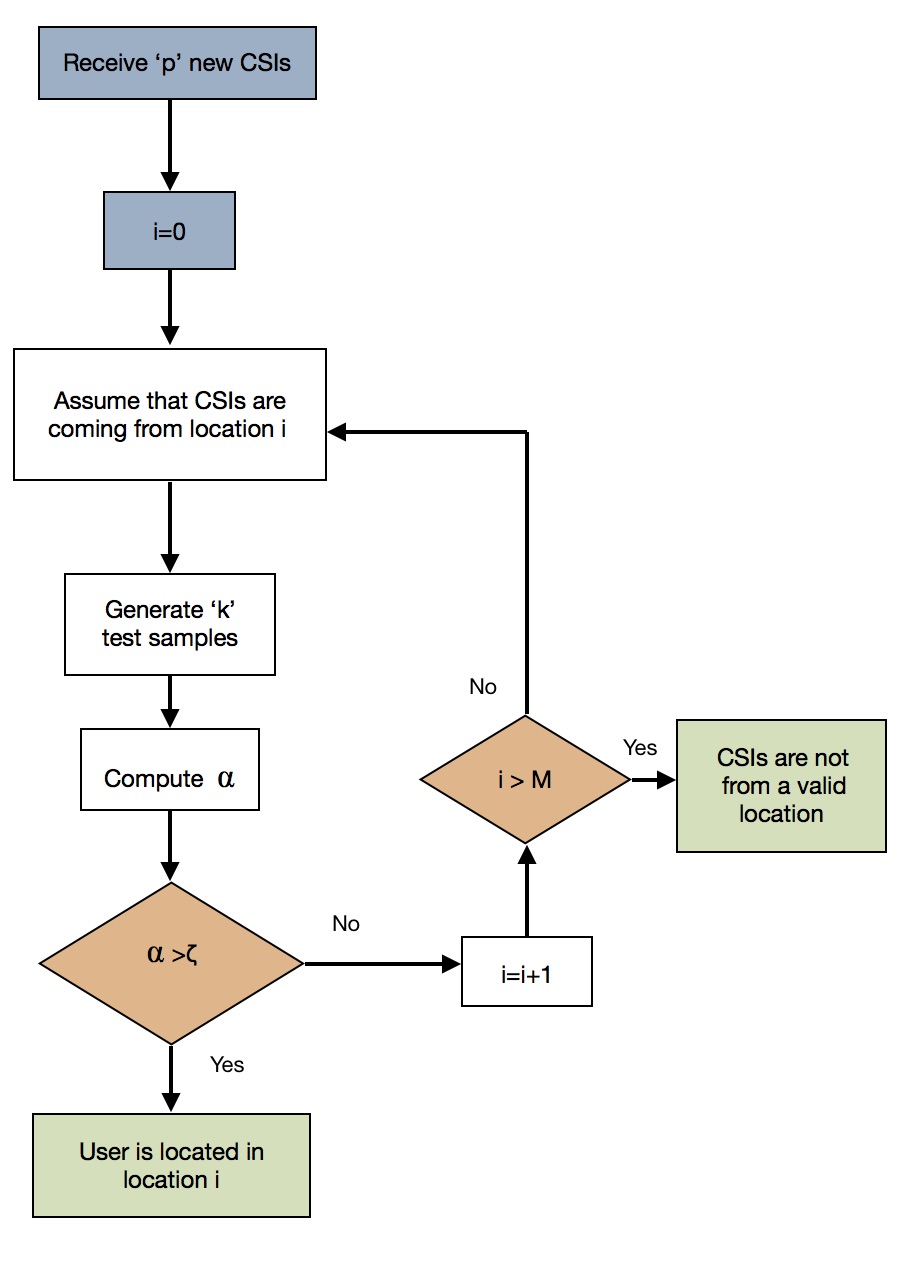}
  \caption{A flowchart of the authentication framework}
  \label{fig:flow1}
\end{figure}

The above procedure should be repeated for all $M$ locations and if none of them pass the majority test the server is not authorizing the user. A flow chart of our framework is shown in figure \ref{fig:flow1}.

\section{Experimental Results}

\begin{figure}
  \centering
  \includegraphics[scale=.15]{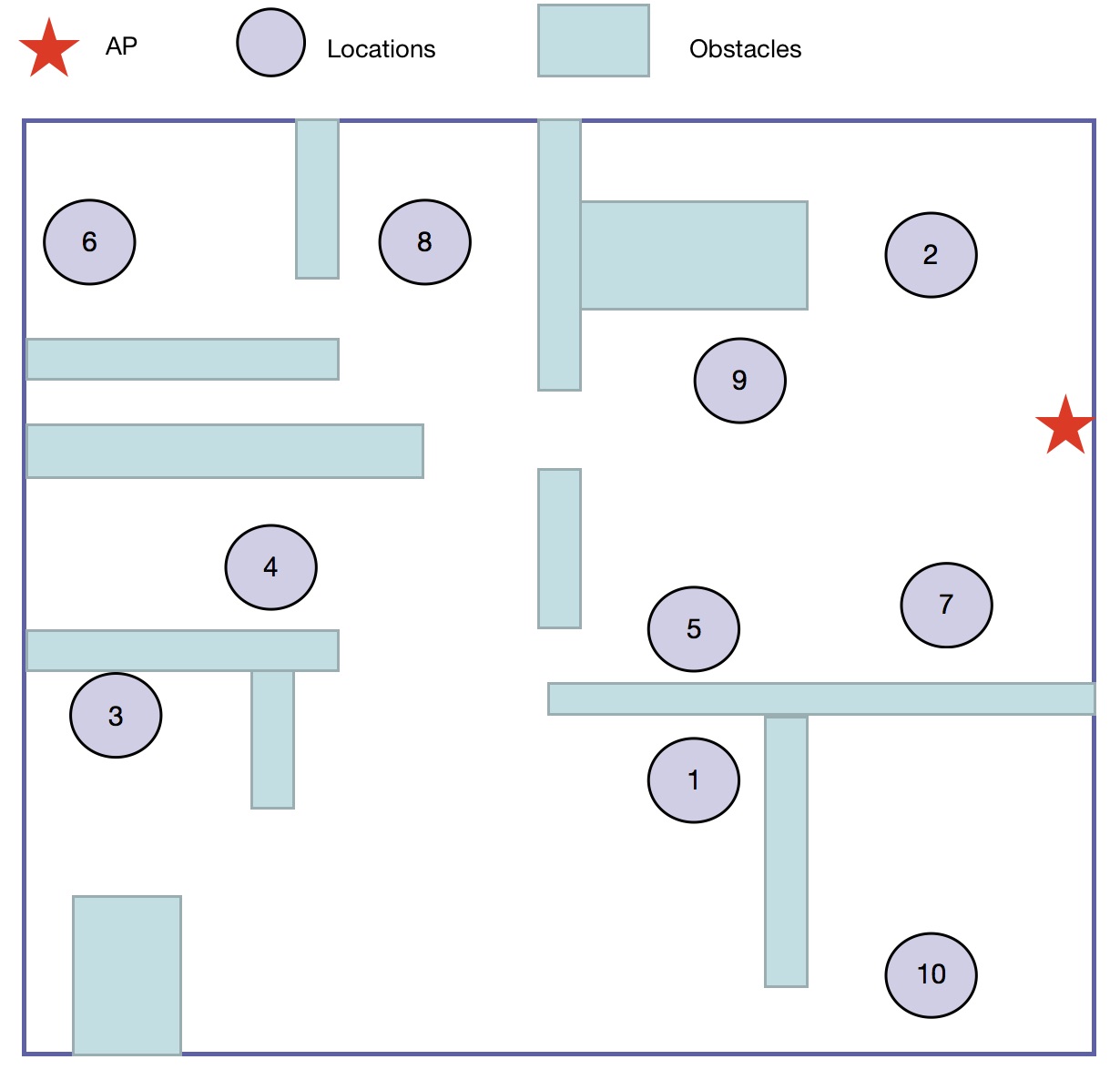}
  \caption{Apartment}
  \label{fig:apartment}
\end{figure}

\begin{figure}
  \centering
  \includegraphics[scale=.15]{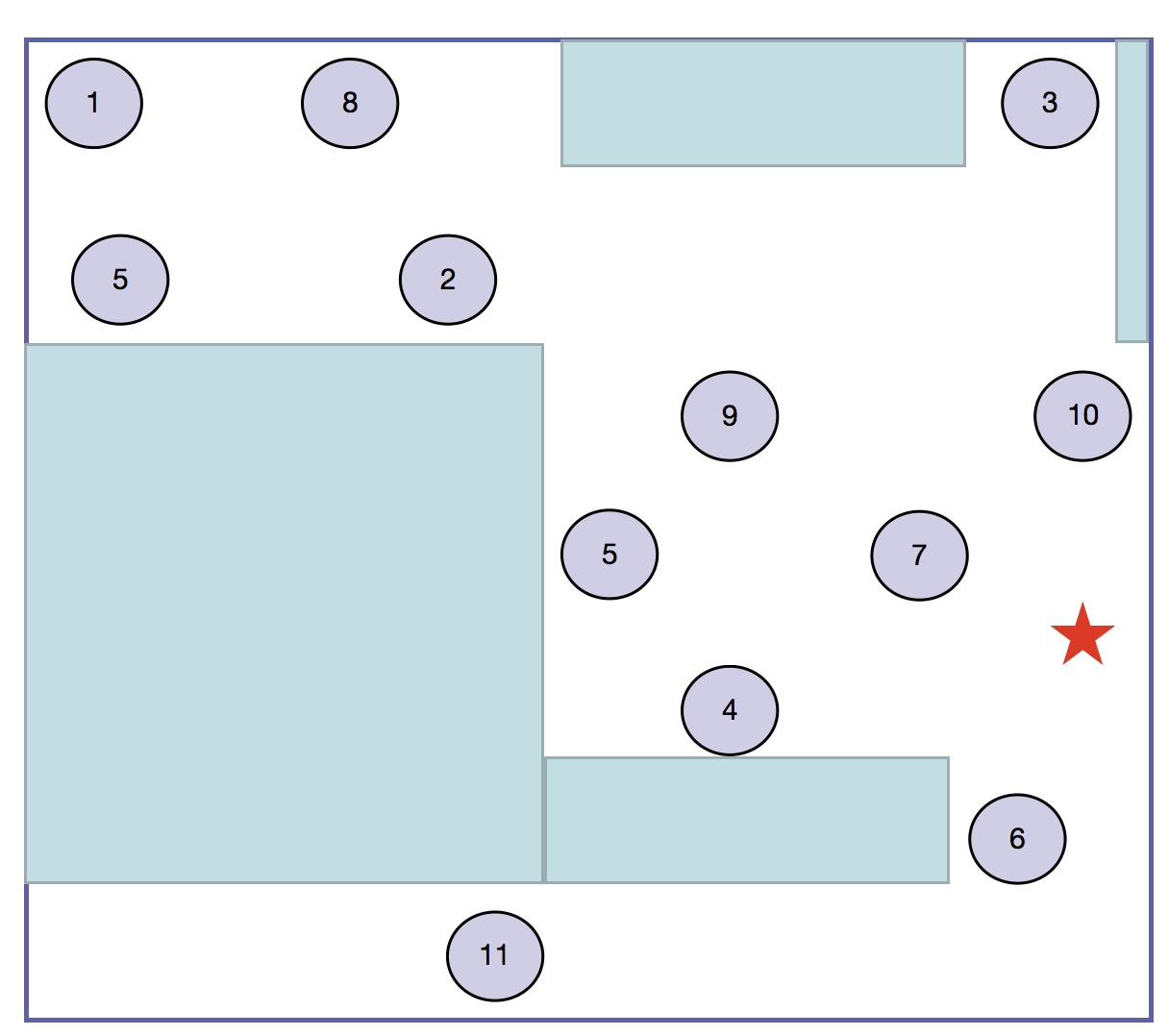}
  \caption{Garage}
  \label{fig:garage}
\end{figure}

We conducted our experiment in two distinct environments: an "apartment" and a "garage and the ramp", their maps are shown in figures \ref{fig:apartment} and \ref{fig:garage}, respectively. In each scenario, we collect CSI data at $Q$ locations out of which we consider $M$ as valid locations. The rest of the locations ($Q-M$) are therefore not valid points. 

The authentication framework thus :
 \begin{itemize}
  \item[1] Need to raise successful authentication flag if we feed it with CSIs from one of the $M$ locations (regardless of the orientation the CSI has been collected).
  \item[2] Should reject authentication if the CSI comes from the rest of the locations.
 \end{itemize}
 The above two criteria are sufficient for making sure only users can be accessed to the network form authenticated locations, but that would also be good if we can find out what is the exact location of the users, i.e., not only saying that the user is asking for a connection from one of the $M$ locations, but also we can exactly determine the location of the user among the $M$ locations.


To collect CSI, in all experiments, we used a Dell laptop as the receiver and a TP-link Wi-Fi router as our AP. The laptop is working on Ubuntu 10.4 LTS (2.6.36 Kernel) OS and an Intel 5300i NIC was implemented on it. Thanks to \cite{ref:csitool}, this NIC has an API using which we can get CSI for each packet it receives. To collect CSI data, the laptop is connected to the AP and should run the ``ping'' command for receiving packets. These packets were sent by the speed of 10 packets per second. The CSI data recorded for each packet is a vector of 30 complex numbers (corresponding to 30 OFDM subcarriers). 

In the following, first we will briefly look into correlation coefficient between the CSI vector after rotation. This results verify that the correlation based scheme would not have satisfactory performance when rotation is happening at the users. The results of the proposed authentication framework  for the two environments will be presented next.

\subsection{CSI Correlation}

\begin{figure}
  \centering
  \includegraphics[scale=.1]{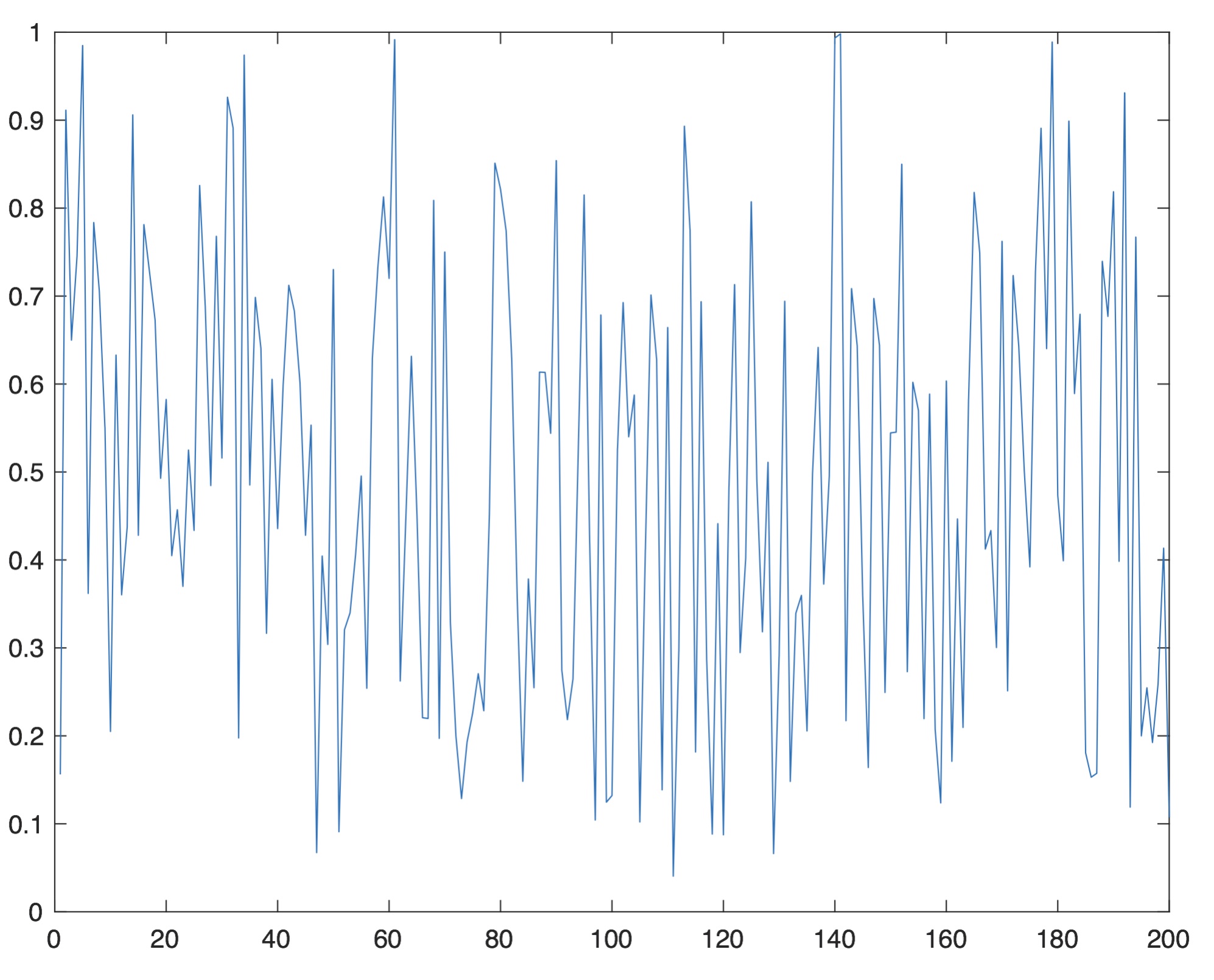}
  \caption{Correlation of subsequent packets of a same position with different angles (position 1).}
  \label{fig:corr}
\end{figure}

\begin{figure}
  \centering
  \includegraphics[scale=.17]{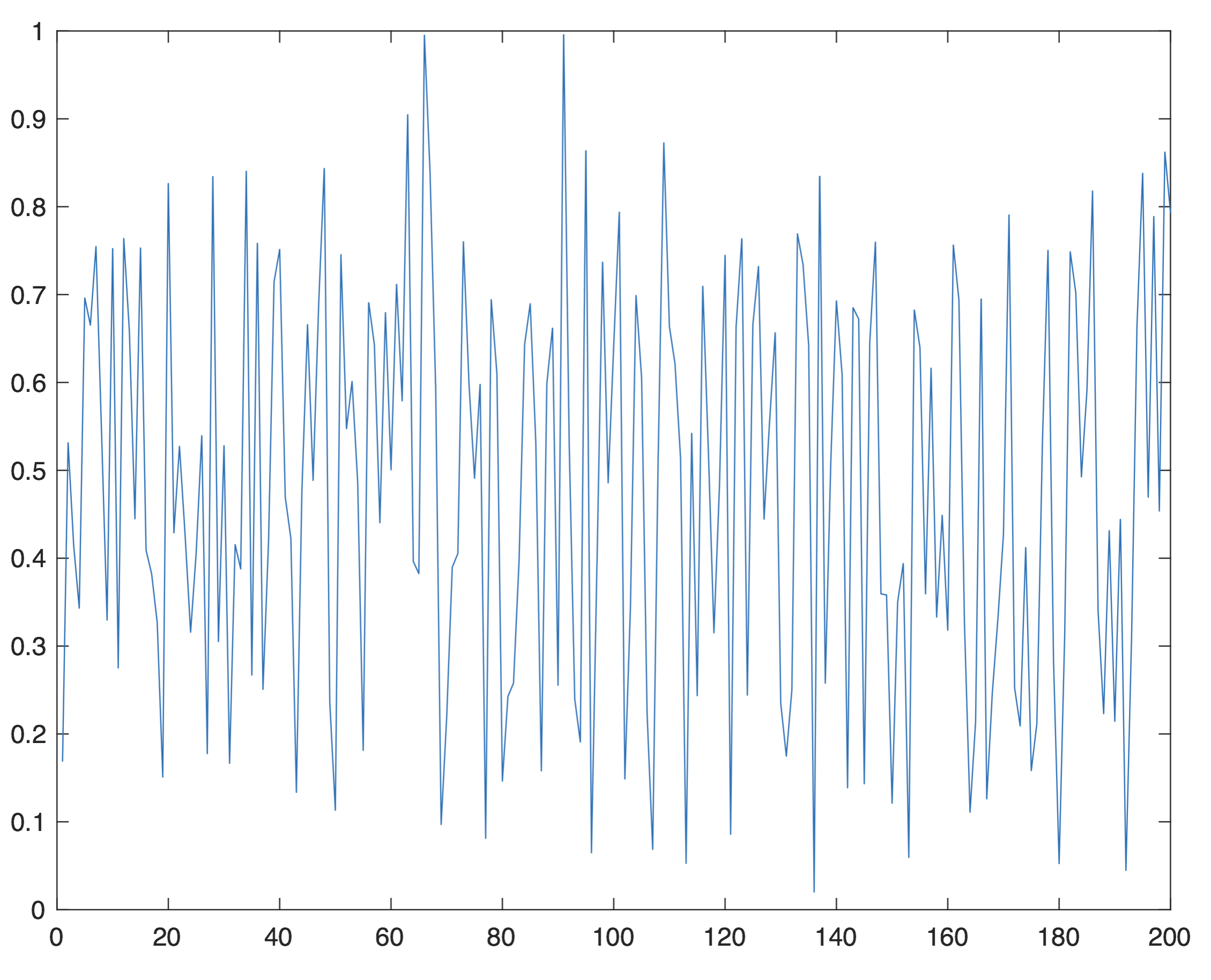}
  \caption{Correlation of subsequent packets of a same position with different angles (position 2).}
  \label{fig:corr2}
\end{figure}

 To see the performance of the CSI correlation, we fixed the laptop's and the AP's position and collected CSI (as described above) for 200 packets while we slowly rotated the device clockwise. For each consecutive CSI measurement vectors of $C_1$ and $C_2$, we calculated the absolute value of correlation coefficient as: 
\begin{equation} \label{eq:7}
\rho=\frac{|C_1^TC_2|}{\|C_1\| \|C_2 \|},
\end{equation}
 where $\|.\|$ is norm and $T$ is the the transposing operation. Here $\rho$ has a value between 0 and 1.  The correlation results for the received 200 packets for two locations are depicted in figure  \ref{fig:corr} and  \ref{fig:corr2}.\\

As it is shown, the correlation between consecutive CSI vectors of a same position with different angles do not show a trend and thus can not be used as a means to identify if the two measurements represent a common position.

\subsection{Authentication's results based on LocNet}

\subsubsection{First Environment (an Apartment)}

 The data for the first experiment was collected in an approximately  $15\times10 m^2$  apartment, Fig. \ref{fig:apartment}. In this environment, we consider 8 (location 1 to 8) as valid locations. At each of these 8 locations, we collected 800 CSI vectors while we slowly rotated the laptop clockwise. The LocNet is then trained using the procedure described in  \ref{ssec:train}.

 To evaluate the model, we should test:
  \begin{itemize}
  \item[I-] How well it can validate the new user if it gets new CSI vectors from location 1 to 8 
  \item[II-] How well it can reject the user if  it gets new CSI vectors from another location (like location 9 and 10 in our environment
  \item[III-] Although not essential for authentication, to test how well we can exactly find out the location of user if it sends the CSI from one of the valid locations (Location 1 to 8), i.e., not only saying that the CSI is coming from a valid point, but also say that which valid location that is. 
 \end{itemize}

\begin{figure}[t]
  \centering
  \includegraphics[scale=.15]{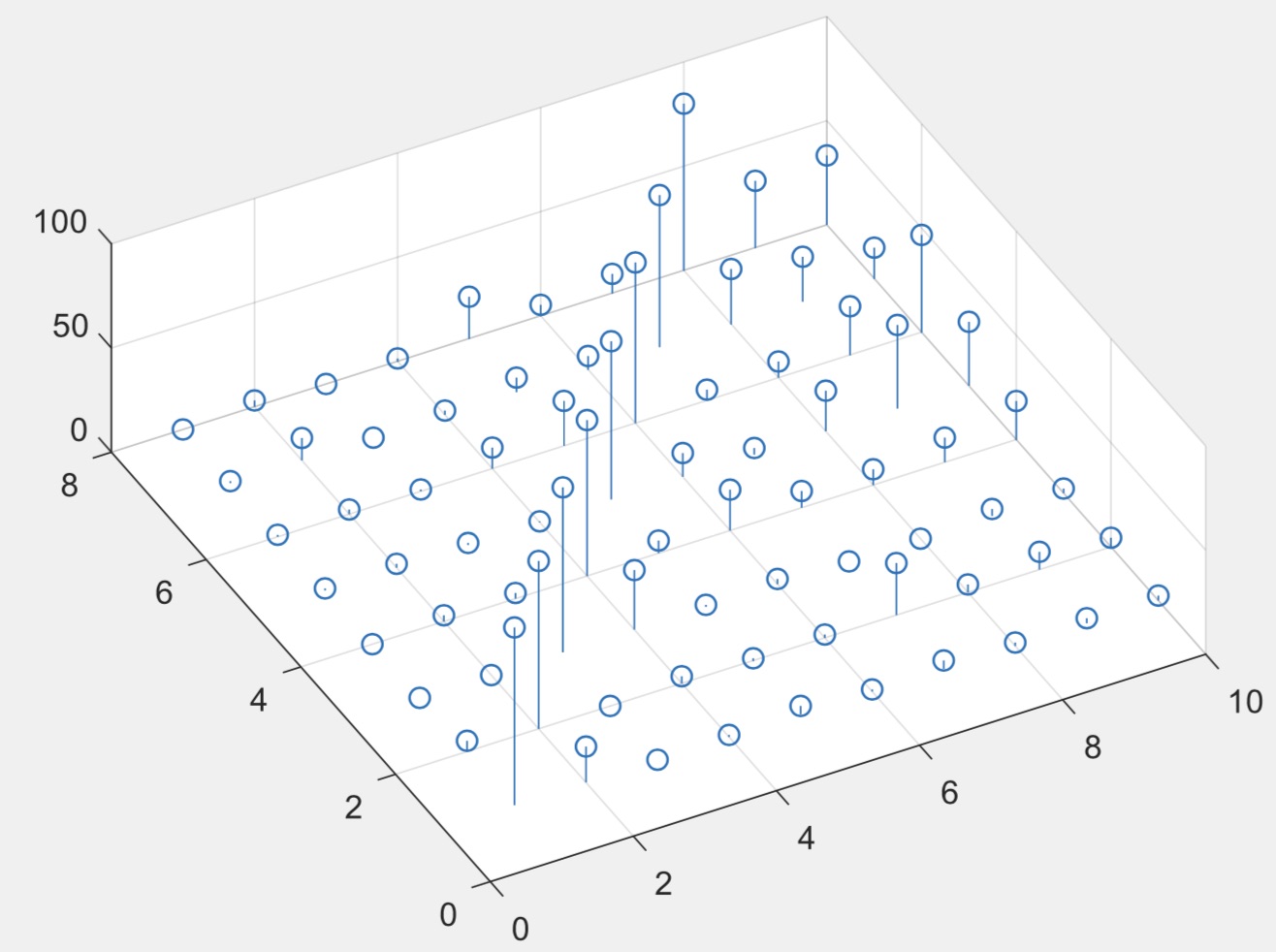}
  \caption{$\alpha\times100$ for the apartment}
  \label{fig:r1}
\end{figure}

 To see the above performance, we collected 100 CSI at all locations (location 1 to 8, and also 9 and 10). The authentication procedure presented in Fig. \ref{fig:flow1} is then used to compute $\alpha$ for each location ($\alpha$ was the ratio between the number of test samples declare that the CSI is from that location and the total number of test samples). The results are shown in table \ref{table:1} and figure \ref{fig:r1}. In table \ref{table:1} , the columns represent the test location that the new CSI is collected, and rows represent the hypothesis location that was used for LocNet, the numbers in the tables shows how often ($r\time100$) the LocNet decide that the test CSI (column) is associated with CSI of the hypothesis location (row). Having these result, and by selecting a value for $\zeta$ the authenticator can decide if the CSI is coming from a valid location or not. 

 As can be seen, in this experiment, by setting $\zeta=0.70$ we can correctly identify the valid locations. Note that the columns related to location 9 and 10 do not have any value larger that $\zeta$ so the authentication server will declare not a valid point upon receiving CSI from these locations.  These results show that the proposed scheme passes test I and II mentioned before. Additionally, as each test location is correctly associated with the correct hypothesis location, the scheme also passes Test III, and it exactly identifies the location of new CSIs.  

%
\subsubsection{Second Environment (a Garage)}
 The second experiment is a garage with its entrance ramp of size approximately $12\times15 m^2$, Fig. \ref{fig:garage}. In this case we consider 9 locations (locations 1 to 9) as valid locations. Similar to the previous case, we collected CSIs at each of these locations while we slowly rotated the laptop clockwise (1800 CSIs) and procedure described in  \ref{ssec:train} used to train the LocNet.

\begin{figure}[t]
  \centering
  \includegraphics[scale=.15]{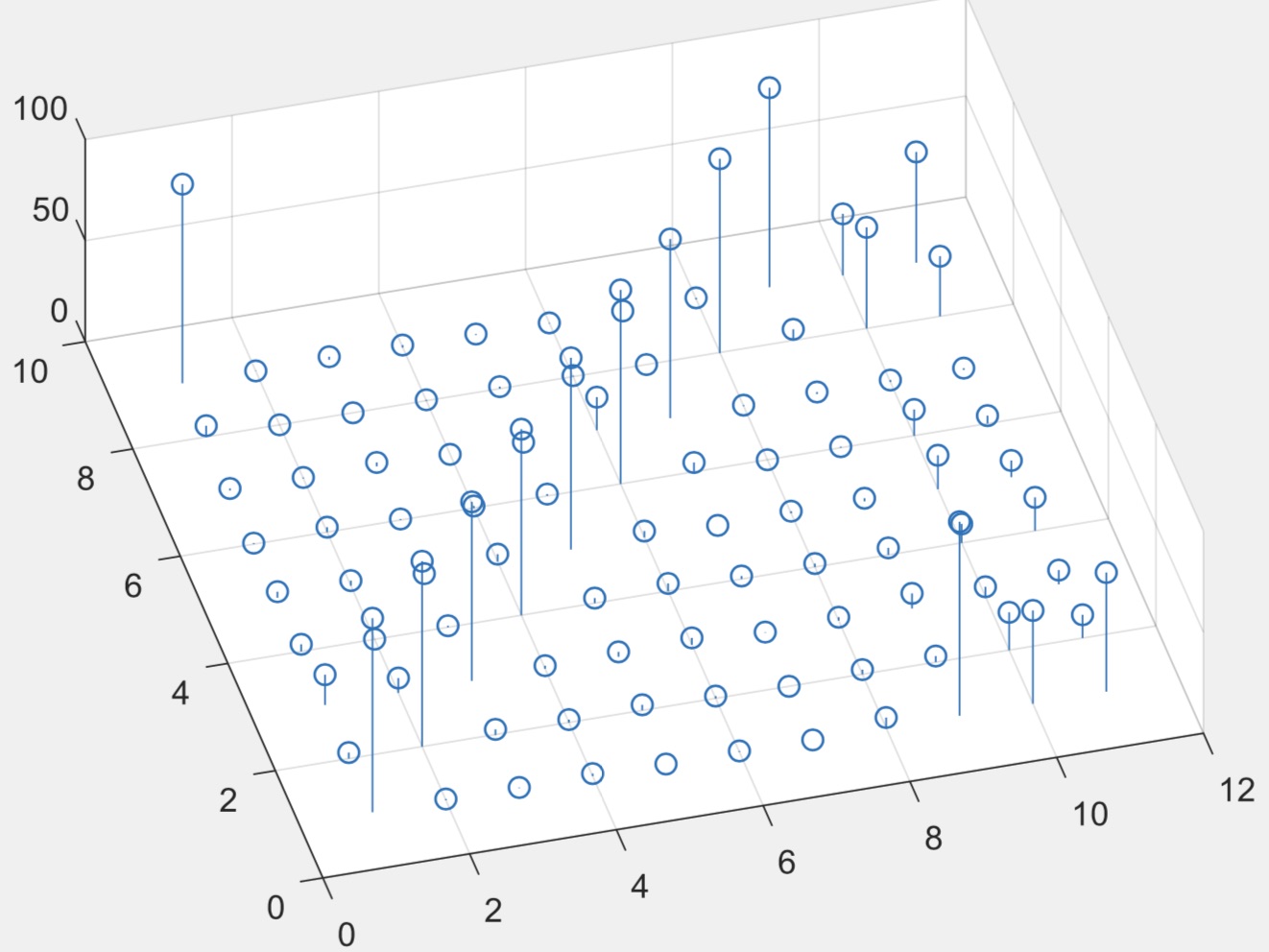}
  \caption{$\alpha\times100$ for the Garage}
  \label{fig:r2}
\end{figure}

 To test the model, we collected 100 new CSI at 11 locations (Valid points: Locations 1 to 9, and two other points which are not authorized to connect). The proposed authentication framework (Fig. \ref{fig:flow1}) is then used to determine if the new CSIs are coming from a valid point or not. The results, showing the values of $r\time100$, are resented in table \ref{table:2} and figure \ref{fig:r2}. As before, the entry $\alpha_{ij}$ of table \ref{table:2} shows the ratio of test cases that the new CSI is actually from Location $j$ and the LocNet estimate that it is coming from  Location $i$. 

 Results of table \ref{table:2} again verify that proposed authentication scheme can successfully differentiate the CSIs coming from valid points and not-valid points. Thus, it passes the two required test cases of Test I and Test II. The results also show that authenticating server is to determine the actual location of the new CSI if it transmitted from location 2 to 8, but it will mixed up location 1 and 9. Note that the confusion about the exact location of the the new CSI is not hurting the authentication accuracy as both of these locations are valid points. However, to eliminate this incorrect estimation, we have tried different network architecture but were not able to resolve it yet. Further investigation is needed on that.

\begin{table}
\centering
\begin{adjustbox}{width=\linewidth, height =15mm}
\small
 \begin{tabular}{||c||c|c| c| c| c| c| c| c| c| c|c| |} 
 \hline
Position & Loc1 & Loc2   & Loc3 & Loc4 & Loc5 & Loc6 & Loc7 & Loc8 & Loc9 & Loc10 \\ 
 \hline\hline
 Loc1 & \textcolor{red}{85.35}   & 17.25 &   0 &        1 &   4 &   1  &  4 &   1.7 &   2.45 &  2.4 \\ 
 \hline
 Loc2 & 5.15 &  \textcolor{red}{80.60} &  0.05  &  3.4  &  1.1  &  1.6 &   25  &  3.8  &  8.55  & 4.4   \\
 \hline
 Loc3 & 0  & 0 &  \textcolor{red}{79.25}  &  28.6   & 1  &  2.5  &  0  &  0  &  3.35  & 2.25 \\
 \hline
 Loc4 & 0  &  2.95 &   2.75 &  \textcolor{red}{74.85}  &  6.1  & 19.6 &   8.05   &      7.55  &  11.85  &  18.5  \\
 \hline
 Loc5 & 1  &  1.9  &  1  &  0.45 &  \textcolor{red}{76}  &  11.4  &  2.9     &    19.5  &  40.15 &  30.75 \\ 
 \hline
 Loc6 &1  & 2.2  &  0.9  &  10.1  &  21.6 &  \textcolor{red}{77.15}  & 5.15  &  7.75     &    23.35  &  46.7\\ 
  \hline
 Loc7 & 0.95  &  10.75 &   0    &     2.1  &  7  & 6.45 &  \textcolor{red}{72.8} &  26.4  &  21.35  &  14.8 \\ 
  \hline
 Loc8 &0  &  3  &  0  &  1.35  &  20.05  &  5.25  &  9.2  & \textcolor{red}{80}  &  32.05 &  33.3   \\ 
  \hline
   \hline
\end{tabular}
\end{adjustbox}
\caption{$\alpha\times100$ for the apartment} 
\label{table:1}
\end{table}

 \begin{table}[t]
\centering
\begin{adjustbox}{width=\linewidth, height =15mm}
\small
 \begin{tabular}{||c||c|c| c| c| c| c| c| c| c| c|c| c||} 
 \hline
Position & Loc1 & Loc2   & Loc3 & Loc4 & Loc5 & Loc6 & Loc7 & Loc8 & Loc9 & Loc10& Loc11 \\ 
 \hline\hline
 Loc1 & \textcolor{red}{96}   & 0.57 &   0.32 &        1.22 &   0 &   0.5  &  0 &   5.1 &  \textcolor{blue}{96}   &  46.15 & 58.95 \\ 
 \hline
 Loc2 & 3 &  \textcolor{red}{91.67} &  2.55  &  1.42  &  2.77  &  1.12 &   0  &  2.27  &  3.02  & 18.7 & 11.52   \\
 \hline
 Loc3 & 15  & 7.37 &  \textcolor{red}{88.57}  &  1.55   & 2.42  &  3.37  &  0.275  &  1.67  &  7.5  & 4.8 & 7.1 \\
 \hline
 Loc4 & 3.5  &  0.07 &   0.77 &  \textcolor{red}{92.02}  &  2.5  & 3.8 &   1.5   &   1.7  &  3.5  &  9.37 & 16.55  \\
 \hline
 Loc5 & 3  &  2.52  &  0  &  3.67 &  \textcolor{red}{94.85}  &  3.17  &  0     &   1.07  &  1.5 &  16.82 & 8.27 \\ 
 \hline
 Loc6 &0.5  & 0.47  &  1.95  &  0  &  0 &  \textcolor{red}{95.95}  & 4.6  & 0     &    0.5  &  12.95 & 3.92\\ 
  \hline
 Loc7 & 1  &  0.47 &   1.95    &     0  &  0  & 16.37 &  \textcolor{red}{88.62} &  0.47  &  1  & 0.97 & 0.9 \\ 
  \hline
 Loc8 &5.25  &  0 &  0  &  0.45  &  1  &  0.5  &  0  & \textcolor{red}{95.82}  &  5.5 &  49.97 & 29.72   \\ 
  \hline
  Loc9 &\textcolor{blue}{98.5}  &  0.15  &  1.2  &  1  &  0.5  & 0.02  &  0  & 0.5  &  \textcolor{red}{98.5} &  30.12 & 54.8   \\ 
  \hline
   \hline
\end{tabular}
\end{adjustbox}
\caption{$\alpha\times100$ for the Garage} 
\label{table:2}
\end{table} 

\section{Conclusions }
 The existing methods for CSI-based authentication use raw Channel State Information (CSI) to determine the authenticity of a user. As the raw CSIs are changing significantly when a user rotates; the existing scheme that are working based on correlation of CSI values do not have high accuracy in these scenarios.
 This paper proposition is to first use raw CSI to find some features which are stable towards user rotation; afterwards, authentication will be done by using these stable features rather than raw CSIs values themselves. The robust CSI features  are derived using deep learning methods. Using the proposed Authentication framework, users will only be able to communicate while they are residing at specific locations and  user rotation won't degrade the authentication accuracy. Experimental results for two scenarios have been reported to verify the performance of the scheme. 


%

%
%
%
%

%

\ifCLASSOPTIONcaptionsoff
  \newpage
\fi

\end{document}